\documentclass[aps,prd,onecolumn,eqsecnum,amsmath,nofootinbib,preprintnumbers]{revtex4}%

\usepackage[dvips]{color,graphicx}
\usepackage{amsfonts,amssymb,theorem,mathrsfs,times}
\usepackage{bm}
{\theorembodyfont{\upshape}
}
{\theorembodyfont{\upshape}
\newtheorem{The}{Theorem}}
{\theorembodyfont{\upshape}
}
{\theorembodyfont{\upshape}
}
{\theorembodyfont{\upshape}
}
{\theorembodyfont{\upshape}
}

\newcommand{\dalm}{\kern1pt\vbox{\hrule height 0.9pt\hbox{\vrule width
0.9pt\hskip 2.5pt\vbox{\vskip 5.5pt}\hskip 3pt\vrule width 0.3pt}\hrule height
0.3pt}\kern1pt}

\begin{document}
\preprint{\hfill {\small {ICTS-USTC-14-12}}}
\title{General proof of (maximum) entropy principle in Lovelock gravity
}

%

\author{Li-Ming Cao$^{a,b}$\footnote[2]{e-mail address:
caolm@ustc.edu.cn}, Jianfei Xu$^{a}$\footnote[1]{e-mail address:
jfxu06@mail.ustc.edu.cn}}


\affiliation{$^a$
Interdisciplinary Center for Theoretical Study\\
University of Science and Technology of China, Hefei, Anhui 230026, China}

\affiliation{$^b$ State Key Laboratory of Theoretical Physics,
Institute of Theoretical Physics, Chinese Academy of Sciences,
P.O. Box 2735, Beijing 100190, China}
%

\date{\today}

\begin{abstract}
We consider a static self-gravitating perfect fluid system in
Lovelock gravity theory. For a spacial region on the hypersurface
orthogonal to static Killing vector, by the Tolman's law of
temperature, the assumption of a fixed total particle number inside
the spacial region, and all of the variations (of relevant fields)
in which the induced metric  and its first derivatives are fixed on
the boundary of the spacial region, then with the help of the
gravitational and fluid equations of the theory, we can prove a
theorem says that the total entropy of  the fluid in this region
takes an extremum value. A converse theorem can also be obtained
following the reverse process of our proof. We also propose the
definition of isolation quasilocally for the system and explain the
physical meaning of the boundary conditions in the proof of the
theorems.
%
\end{abstract}


\maketitle


\section{Introduction}
Black holes are fundamental objects in gravity theory which has been
studied for a long time. Several breakthrough developments had been
achieved about forty years ago. At the beginning of the 1970's,  it
is found that the laws of the mechanics of black holes  are very
similar to the usual four laws of thermodynamics~\cite{bardeen1973}.
Soon after, by studying the quantum effects of scalar field around a
black hole,  Hawking found that the black hole behaves like a
blackbody with a temperature which is proportional to its surface
gravity~\cite{hawking}, and then the earlier proposal of the entropy
of the black hole by Bekenstein could be confirmed to be one quarter
of the horizon area~\cite{Bekenstein}. Due to this celebrated work,
the black hole mechanics is promoted to black hole thermodynamics.
Since that time black hole thermodynamics has drawn a lot of
attention, and has been widely studied during the past decades
because people believe it might  be a window in which one can catch
sight of some important fundamental theories, such as the so-called
quantum gravity theory.

Roughly speaking, there are two ways to approach the thermodynamics
of black holes depending on the definitions of the thermodynamic
quantities for associated spacetimes. Because of the equivalence
principle in general relativity, some definitions of density for
usual matter, such as energy density and entropy density are not
valid for gravitational field. At most, we can define the energy of
the gravitational field quasilocally. In the traditional
construction of the thermodynamics of black holes, the thermodynamic
quantities are identified to the global quantities defined at the
infinities of the spacetimes, such as ADM mass, angular momentum,
and charges of gauge fields. Fruitful results have been obtained
along this way; for example, the thermodynamics of stationary black
holes in general diffeomorphism invariant gravity theory has been
established~\cite{Wald:1999vt} (and references therein). However, in
general, it is quite difficult to extract some useful local
information of the spacetimes from this kind of thermodynamics of
black holes.  In 1993, Brown and York walked along another way and
developed a new method to define the thermodynamic quantities
quasilocally by using of a natural generalization of Hamilton-Jacobi
analysis of action functional~\cite{Brown:1992br}. New notions on
the definition of the horizons of black holes were also proposed
soon after by some researchers~\cite{Hayward:1994bu,
Ashtekar:2004cn}. Since then, black hole thermodynamics can be
studied quasilocally, and the gravitational equations can be
obtained from these quasilocal thermodynamic quantities and
associated thermodynamic relations with some additional assumptions.
This quasilocal approach allows us to turn the logic around and
study the gravitational equations from the laws of thermodynamics.
Actually, Jacobson has shown that Einstein equations are the state
equation which can be derived from the Clausius relation by using of
local Rindler horizon and Unruh temperature~\cite{Jac,Jac1}. See
also~\cite{c1} and related works on some quasilocal horizons in
dynamical spacetimes with spherically symmetry. These remarkable
works inspire us to believe that the gravity  and thermodynamics
should have some deep and profound connection.

The thermodynamics of black holes heavily depends the quantum field
theory in curved spacetimes. Quantum effects allow us to regard
black holes as  real thermodynamic systems. However, besides the
black holes, there are also a lot of  self-gravitating systems
without horizons in general relativity and other possible gravity
theories. Of course, the thermodynamics of these self-gravitating
systems is very different from the thermodynamics of black holes.
For example, the Hawking temperature does not exist in these
systems. According to the work of Jacobson, we know that the
gravitational equations can be deduced from the laws of
thermodynamics. (This work is based on the thermodynamics of the
local Rinder horizon, and some results from the quantum field in
curved spacetimes have been used, such as the Unruh effect.) So a
question naturally arises - whether it is possible to get the
gravitational equations from the thermodynamics of the usual matter
fields living in the curved spacetimes? Although there are no
horizons and black holes in these cases, the gravitational equations
and the laws of thermodynamics govern the same thing, i.e., the
distribution of the matter fields (or equilibrium state of the
matter fields) in stationary spacetimes. So  some equivalent
description among gravitational equations and thermodynamic laws
might exist for these self-gravitating systems. For a spherical
radiation system, Sorkin, Wald, and Zhang (SWZ) have shown that one
can deduce the Tolman-Oppenheimer-Volkoff (TOV) equation from
Hamiltonian constraint when the total entropy of radiation is in
extremum~\cite{wald}. Gao generalized SWZ's work to an arbitrary
perfect fluid in static spherical spacetime and successfully got the
TOV equation for this fluid ~\cite{gaosijie}. Recently, a more
general proof of the (maximum) entropy principle in the case of
static spacetime without the spherical symmetry has been completed
in~\cite{Anastopoulos:2013xdk, Fang:2013oka}.

However, all of the above discussions are limited in Einstein
gravity. One can ask whether the entropy principle is still valid or
not in other gravity theories. Lovelock gravity theory is a natural
generalization of Einstein gravity to higher dimensions. The action
of this theory includes higher derivative terms with respect to
metric, while the equations of motion still keep the derivatives up
to second order~\cite{Love}. Due to the development of supergravity
and string theory, Locklock gravity theory becomes more and more
important. For example, people find that the higher order Lovelock
terms appear in the higher order $\alpha'$ expansion of string
amplitude~\cite{Zwie,Lowenergylimit,Lowenergylimit1,Lowenergylimit2,Lowenergylimit3}.
Thus it is interesting to discuss the (maximum) entropy principle in
this generalized gravity theory. The (maximum) entropy principle for
the self-gravitating perfect fluid in an $n$-dimensional Lovelock
gravity with the symmetry of an $(n-2)$-dimensional maximally
symmetric space has been studied by the present authors
in~\cite{Cao:2013xy}, where the generalized TOV equations have been
derived from both gravity field equations and the (maximum) entropy
principle of perfect fluid.

In this paper, we will present a  proof  of a theorem that the
(maximum) entropy principle in the Lovelock gravity theory generally
holds  without considering the symmetry of an $(n-2)$-dimensional
maximally symmetric space. The only symmetry we will consider is the
static condition of the self-gravitating system. Assuming that the
Tolman's law of the temperature holds for the perfect fluid  in
curved spacetimes,  and imposing some boundary conditions in the
variations of relevant fields, we show that the entropy of the fluid
inside an $(n-2)$-dimensional spacelike surface [which is embedded
in an $(n-1)$-dimensional hypersurface orthogonal to the static
Killing vector of the spacetime ] takes extremum value. Our
discussion is focused on the system inside the $(n-2)$-dimensional
spacelike surface of the static spacetimes, so it is not hard to
extract some local information, i.e., a part of gravitational
equations, from the law of thermodynamics of the system. This
suggests that the converse theorem can  be read out from our proof.
Related work on the thermodynamics of self-gravitating system can be
found in Ref.~\cite{Green:2013ica} in which global quantities, such
as ADM mass, have been used to discuss the thermodynamic stability
of the system.

We also study the physical meaning of boundary conditions. We found
that the boundary conditions which seem nontransparent in the proof
of the theorems finally turn out to be the isolation condition which
is necessary of applicability of (maximum) entropy principle. Since
the system we consider here is a quasilocal system, so the isolation
here is quasilocally defined. In this sense, our proof is
self-contained.

This paper is organized as follows: In Sec.II, for static spacetime,
we study the equations of motion of the so-called
Einstien-Gauss-Bonnet gravity which can be viewed as a special case
of the Lovelock gravity up to second order. In Sec.III, we study the
thermodynamics of the perfect fluid in the Einstien-Gauss-Bonnet
theory and prove a theorem which relates the gravitational equations
and the (maximum) entropy principle of the fluid. In Sec.IV, we
generalize our proof to the general Lovelock gravity theory. In
Sec.V, we check our previous work in which we studied the entropy
principle in Lovelock gravity with an $(n-2)$-dimensional maximally
symmetric space by using the present method. In Sec.VI, in contrast
to the isolation condition of the usual thermodynamic system, we
present the definition of an isolated system quasilocally. The last
section is devoted to some conclusions and discussion.
\section{The equations of motion of Einstein-Gauss-Bonnet Gravity in static spacetimes}
\label{sec:II} The Einstein-Gauss-Bonnet gravity is a typical
example of the general Lovelock gravity. As a warm-up and an
example, in this section, we consider the case of the
Einstein-Gauss-Bonnet gravity in an $n$-dimensional spacetime $(M,
g_{ab})$. The action of this system can be written as
\begin{equation}
I=\frac{1}{2}\int \epsilon \big(\mathscr{R}+\alpha~
\mathscr{L}_{GB}\big)+ I_{\mathrm{matter}}\, ,
\end{equation}
where $\alpha$ is the so-called Gauss-Bonnet coupling constant,
$\epsilon$ is the natural volume element associated with the metric
$g_{ab}$, and $\mathscr{L}_{GB}$ is the Gauss-Bonnet term which has
a form
\begin{equation}
\mathscr{L}_{GB}=\mathscr{R}^2-4\mathscr{R}_{ab}\mathscr{R}^{ab}+\mathscr{R}_{abcd}\mathscr{R}^{abcd}\,
.
\end{equation}
Here, we have used curly alphabet to denote the geometric quantities in the $n$-dimension. For instance, $\mathscr{R}_{abcd}$, $\mathscr{R}_{ab}$, and $\mathscr{R}$ are
the curvature tensor, Ricci  tensor, and scalar curvature  for the $n$-dimensional spacetime respectively. The symbol $I_{\mathrm{matter}}$ represents the action for
the matter fields. The variation of the action with respect to the metric yields the gravitational equation
\begin{equation}
\label{EOMGB}
\mathscr{G}_{ab}+\alpha\mathscr{H}_{ab}=T_{ab}\, ,
\end{equation}
where $\mathscr{G}_{ab}$ is the Einstein tensor of the spacetime $(M, g_{ab})$ and $\mathscr{H}_{ab}$ is given by
\begin{eqnarray}
\mathscr{H}_{ab}= 2 \mathscr{R}_{a c d e} \mathscr{R}_{b}{}^{c d e}  - 4\mathscr{R}^{c d} \mathscr{R}_{a c b d}
- 4 \mathscr{R}_{b c} \mathscr{R}_{a}{}^{c}+ 2\mathscr{R} \mathscr{R}_{a b}-\frac{1}{2} \mathscr{L}_{GB}~g_{ab}\, .
\end{eqnarray}
Since we have set $8\pi G=1$, the right-hand side of
Eq.(\ref{EOMGB}) is simply the energy-momentum tensor $T_{ab}$ for
the matter fields.

The spacetime $(M, g_{ab})$ we are considering is  assumed to be stationary. This suggests that we have a timelike Killing vector field
$K^a$, i.e., $K^a$ satisfies Killing equation $\nabla_{(a}K_{b)}=0$,
where $\nabla_a$ is the covariant derivative  compatible
with the metric $g_{ab}$(we use the notations and conventions in~\cite{Wald:1984rg}). Following the notation by Geroch~\cite{Geroch:1970nt}, in the
region where $\lambda=K^aK_a\ne 0$, we can define a metric on the
orbit space $\Sigma$ of the Killing field
\begin{equation}
\label{inducedmetric}
h_{ab}=g_{ab}-\lambda^{-1}K_aK_b\, .
\end{equation}
If the Killing vector field is hypersurface orthogonal, i.e., Frobenius condition
\begin{equation}
\label{twist}
K_{[a}\nabla_{b}K_{c]}=0
\end{equation}
is satisfied, the orbit space $\Sigma$ can be viewed as a
hypersurface embedded in the spacetime.  We always assume this condition is satisfied in following discussion. In other words,
the spacetime is further assumed to be static.
Considering the Frobenius condition (\ref{twist}), it is not hard to find
\begin{eqnarray}
\label{RabcdtoRabcd}
R_{abcd}=\mathscr{R}_{abcd}-2\lambda^{-2}{K}_{[a}{\nabla}_{b]}{{\nabla}_{[c}{\lambda} }  {K}_{d]}
+  \lambda^{-3} {K}_{[a} {\nabla}_{b]}{\lambda} {\nabla}_{[c}{\lambda}   {K}_{d]}
   \, .
\end{eqnarray}
Here, $R_{abcd}$ is the intrinsic curvature of the hypersurface
$(\Sigma, h_{ab})$. Based on the relation, we have the following
decompositions
\begin{eqnarray}
\mathscr{R}
&=&R-\frac{1}{2}\lambda^{-1}h_{ab}\big(2D^aD^b\lambda-\lambda^{-1}D^a\lambda
D^b\lambda\big)\, ,
\end{eqnarray}
and
\begin{eqnarray}
\mathscr{L}_{GB}=L_{GB}+2\lambda^{-1}
G_{ab}\big(2D^aD^b\lambda-\lambda^{-1}D^a\lambda D^b\lambda\big)\, ,
\end{eqnarray}
where $D_a$ is the covariant derivative operator which is compatible with
the induced metric $h_{ab}$ in Eq.(\ref{inducedmetric}), and $R$, $L_{GB}$,
and $G_{ab}$ are the scalar curvature, Gauss-Bonnet term,  and,
Einstein tensor for the hypersurface $(\Sigma, h_{ab})$ respectively. Furthermore, we have
\begin{eqnarray}
\mathscr{G}_{ab}=G_{ab}-\frac{1}{2}\lambda^{-1}RK_{a}K_{b}+\frac{1}{2}h_{a[b}h_{d]c}\big(2\lambda^{-1}D^cD^d\lambda-\lambda^{-2}D^c\lambda
D^d\lambda\big)\, ,
\end{eqnarray}
and
\begin{eqnarray}
\mathscr{H}_{ab}&=& H_{ab}-\frac{1}{2}\lambda^{-1} L_{GB}K_aK_b +
\Big[R_{a c b d} -2 R_{a [b}  h_{d]c}
+ 2R_{c[b}  h_{d]a}\nonumber\\
&+&R
h_{a[b}h_{d]c}\Big]\big(2\lambda^{-1}D^cD^d\lambda-\lambda^{-2}D^c\lambda
D^d\lambda\big)\, ,
\end{eqnarray}
with
\begin{eqnarray}
H_{ab}= 2 R_{a c d e} R_{b}{}^{c d e}  - 4R^{c d} R_{a c b d} - 4 R_{b c} R_{a}{}^{c}+ 2R R_{a b}-\frac{1}{2} L_{GB}~h_{ab}\, .
\end{eqnarray}
From the above equations, it is easy to find following relations
\begin{equation}\label{energy1}
\lambda^{-1}\mathscr{G}_{ab}K^aK^b=-\frac{1}{2}R\, ,
\end{equation}
and
\begin{equation}\label{energy2}
\lambda^{-1}\mathscr{H}_{ab}K^aK^b=-\frac{1}{2}L_{GB}\, ,
\end{equation}
and
\begin{equation}\label{pressure1}
h_a{}^ch_b{}^d\mathscr{G}_{cd}=
G_{ab}+\frac{1}{2}h_{a[b}h_{d]c}\big(2\lambda^{-1}D^cD^d\lambda-\lambda^{-2}D^c\lambda
D^d\lambda\big)\, ,
\end{equation}
and
\begin{eqnarray}\label{pressure2}
h_a{}^ch_b{}^d\mathscr{H}_{cd}= H_{ab} + \Big[R_{a c b d} -2 R_{a
[b} h_{c]d} + 2R_{c[b}  h_{d]a}+R
h_{a[b}h_{d]c}\Big]\big(2\lambda^{-1}D^cD^d\lambda-\lambda^{-2}D^c\lambda
D^d\lambda\big)\, .
\end{eqnarray}
These relations are important in our proof of the entropy theorem in the next section.

\section{Self-gravitating Fluid in Static Spacetime and (Maximum) Entropy Principle}
In this section,  we are going to analyze the self-gravitating fluid
in the Einstein-Gauss-Bonnet gravity. In the static spacetime
$(M,g_{ab})$, we assume that the Tolman's law holds, which says that
the local temperature $T$ of the fluid satisfies
\begin{equation}\label{Tol}
T\sqrt{-\lambda}=T_0\, ,
\end{equation}
where  $T_0$ is a
constant which can be viewed as the local temperature of the fluid at  some reference points with $\lambda=-1$. This relation is essential in the construction of an equilibrium state matter distribution in a curved spacetime, and it is popular satisfied in general stationary systems.  Without loss of generality we shall
take $T_0=1$.

Now, we can present a theorem which relates the (maximum) entropy
principle of the fluid and the equations of motion in this static spacetime listed in the previous section.

\begin{The}
\label{theorem1} - Consider a self-gravitating perfect fluid in a
static $n$-dimensional spacetime $(M,g_{ab})$ in
Einstein-Gauss-Bonnet gravity and $\Sigma$ is an $(n-1)$-dimensional
hypersurface orthogonal to the static Killing vector.
Let $C$ be a region inside $\Sigma$ with a boundary $\partial{C}$
and $h_{ab}$ be the induced metric on $\Sigma$. Assume that the
temperature of the fluid obeys Tolman's law and equations of motion
of both gravity and fluid are satisfied in $C$. Then the fluid is
distributed such that its total entropy in $C$ is an extremum for
fixed total particle number in $C$ and for all the variations in
which $h_{ab}$ and its first derivatives are fixed on $\partial{C}$.
\end{The}

\textit{Proof. -} The integral curves of the static Killing vector
field $K^a$ can be viewed as some static observers in the spacetime,
and the velocity vector field of such observers are given by
\begin{equation}
u^a=\frac{K^a}{\sqrt{-\lambda}}\, .
\end{equation}
Obviously, $u^a$ is just the unit norm of the hypersurface $\Sigma$. The acceleration vector associated with these observers has a form
\begin{equation}
A_a=\frac{\nabla_a\lambda}{2\lambda}\, .
\end{equation}
For a general perfect fluid as discussed in~\cite{Cao:2013xy}, the
energy-momentum tensor $T_{ab}$ takes a form
\begin{equation}
T_{ab}=\rho u_au_b+ph_{ab}\, ,
\end{equation}
where $\rho$ and $p$ are the energy density and pressure of the fluid
respectively. In another word, $u^a$ is also the velocity of the comoving observers of the fluid.
The entropy density $s$ is taken to be a function of
the energy density $\rho$ and particle number density $n$ (do not confuse with the dimension of the spacetime), i.e.,
$s=s(\rho,n)$. The standard first law of thermodynamics in terms of these densities
and Gibbs-Duhem relation are listed here as follows
\begin{equation}\label{fl}
d\rho=Tds+\mu dn\, ,
\end{equation}
\begin{equation}\label{GD}
s=\frac{1}{T}(\rho+p-\mu n)\, ,
\end{equation}
where $\mu$ is the chemical potential conjugating to the particle number density $n$. It can be shown from
conservation law $\nabla_aT^{ab}=0$ and static conditions one gets
\begin{equation}
\nabla_ap=-(\rho+p)A_a=-(\rho+p)\frac{\nabla_a\lambda}{2\lambda}\, ,
\end{equation}
together with Eqs.$(\ref{Tol})$, $(\ref{fl})$, and $(\ref{GD})$, we
find
\begin{equation}
\frac{\nabla_a\mu}{\mu}=-\frac{\nabla_a\lambda}{2\lambda}\, ,
\end{equation}
which leads to
\begin{equation}
\mu\sqrt{-\lambda}=\mathrm{constant}\, ,
\end{equation}
or
\begin{equation}\label{const}
\frac{\mu}{T}=\mathrm{constant}\, .
\end{equation}
The total entropy $S$ inside the region $C$ on $\Sigma$ is defined as the integral of the entropy density, i.e.,
\begin{equation}
S=\int_C\bar{\epsilon}~s(\rho,n)\, ,
\end{equation}
where $\bar{\epsilon}$ is the  volume element of $\Sigma$ associated with the induced metric $h_{ab}$, and
invoking the local first law of thermodynamics (\ref{fl}), the
variation of total entropy yields
\begin{eqnarray}
\delta
S&=&\int_C\Big[s~\delta\bar{\epsilon}+\bar{\epsilon}~\Big(\frac{\partial
s}{\partial\rho}\delta\rho+\frac{\partial s}{\partial n}\delta
n\Big)\Big]\nonumber\\
&=&\int_C\Big[s~\delta\bar{\epsilon}+\bar{\epsilon}~\Big(\frac{1}{T}\delta\rho-\frac{\mu}{T}\delta
n\Big)\Big]\, .
\end{eqnarray}
Similarly, the total number of particle $N$ is an integral
\begin{equation}
N=\int_C\bar{\epsilon}~n\, .
\end{equation}
So the fixed total particle number in $C$ yields the following
constraint
\begin{equation}
\int_C \bar{\epsilon}~\delta n=-\int_C n~\delta\bar{\epsilon}\, .
\end{equation}
With this constraint, the variation of the total entropy becomes
\begin{eqnarray}\label{vS}
\delta
S&=&\int_C\bar{\epsilon}~\Big(\frac{1}{2}\frac{\rho+p}{T}h^{ab}\delta
h_{ab}+\frac{1}{T}\delta\rho\Big)\nonumber\\
&=&\int_C\Big[\bar{\epsilon}~\frac{p}{2T}h^{ab}\delta
h_{ab}+\frac{1}{T}\delta(\bar{\epsilon}~\rho)\Big]\, ,
\end{eqnarray}
where we have used $\mu/T=\mathrm{constant}$ and
$\delta\bar{\epsilon}=(1/2)~\bar{\epsilon}~h^{ab}\delta h_{ab}$. The
variations we perform here are only restricted on spacelike
hypersurface $\Sigma$.

For the perfect fluid in an equilibrium state, the total entropy
must take maximal value according to the maximum entropy principle.
So our purpose is to proof the extremum condition $\delta S=0$ from
the gravitational equations which has been studied in
sec.\ref{Sec:II}.

From Eqs. (\ref{energy1}) and (\ref{energy2}), one can easily get
the Hamiltonian constraint of the theory
\begin{equation}
\label{EC}
\rho=\frac{1}{2}\big(R+\alpha L_{GB}\big)\, ,
\end{equation}
while Eqs.(\ref{pressure1}) and (\ref{pressure2})  lead to evolution
equations
\begin{eqnarray}\label{MC}
ph^{ab}&=&G^{ab}+\alpha H^{ab}
+\frac{1}{2}h^{a[b}h^{d]c}\big(2\lambda^{-1}D_cD_d\lambda-\lambda^{-2}D_c\lambda
D_d\lambda\big)+\alpha\Big[R^{a c b d} -2 R^{a [b}
h^{c]d}\nonumber\\
&+& 2R^{c[b}  h^{d]a}+R
h^{a[b}h^{d]c}\Big]\big(2\lambda^{-1}D_cD_d\lambda-\lambda^{-2}D_c\lambda
D_d\lambda\big)\, .
\end{eqnarray}
We can calculate the second term on the right-hand side of
Eq.(\ref{vS}) by using the Hamiltonian constraint (\ref{EC}).
Actually, we have
\begin{equation}\label{vE}
\int_C\frac{1}{T}\delta\big(\bar{\epsilon}~\rho\big)=\frac{1}{2}\int_C\frac{1}{T}\delta\Big[\bar{\epsilon}~(R+\alpha
L_{GB})\Big]=\frac{1}{2}\int_C\bar{\epsilon}~\frac{1}{T}\Big[-\big(G^{ab}+\alpha
H^{ab}\big)\delta h_{ab}+\mathcal{B}_1+\alpha\mathcal{B}_2\Big]\, ,
\end{equation}
where $\mathcal{B}_1$ and $\mathcal{B}_2$ are total derivative terms
which come from the variation of the Einstein-Hilbert term and
Gauss-Bonnet term in the right-hand side of Eq.(\ref{EC})
respectively. Explicitly, they  are given by
\begin{equation}
\mathcal{B}_1=D^aD^b\delta h_{ab}-D^a(h^{bc}D_a\delta h_{bc})\, ,
\end{equation}
and
\begin{eqnarray}
&&\mathcal{B}_2=2D^a\big(RD^b\delta h_{ab}\big)-2D^b\big(D^aR\delta
h_{ab}\big)\nonumber\\
&&-2D^a\big(Rh^{bc}D_a\delta
h_{bc}\big)+2D_a\big(D^aRh^{bc}\delta h_{bc}\big)\nonumber\\
&&-8D_b\big(R^{ac}h^{bd}D_a\delta
h_{cd}\big)+8D_a\big(D_bR^{ac}h^{bd}\delta
h_{cd}\big)\nonumber\\
&&+4D_b\big(R^{ac}h^{bd}D_d\delta
h_{ac}\big)-4D_d\big(D_bR^{ac}h^{bd}\delta h_{ad}\big)\nonumber\\
&&+4D_a\big(R^{ac}h^{bd}D_c\delta
h_{bd}\big)-4D_c\big(D_aR^{ac}h^{bd}\delta
h_{bd}\big)\nonumber\\
&&+2D_b\big(R^{abcd}D_c\delta
h_{ad}\big)-2D_c\big(D_bR^{abcd}\delta h_{ad}\big)\nonumber\\
&&-2D_b\big(R^{abcd}D_d\delta h_{ac}\big)+2D_d\big(D_bR^{abcd}\delta
h_{ac}\big)\, .
\end{eqnarray}
By using integration by parts, and dropping the surface terms with fixed $h_{ab}$
and its first derivative, one can rewrite the last two terms of
Eq.(\ref{vE}) as  follows
\begin{eqnarray}\label{vE1}
&&\frac{1}{2}\int_C\bar{\epsilon}~\frac{\mathcal{B}_1}{T}=\frac{1}{2}\int_C\bar{\epsilon}~\frac{1}{T}\Big[D^aD^b\delta
h_{ab}-D^a(h^{bc}D_a\delta h_{bc})\Big]\nonumber\\
&&=\frac{1}{2}\int_C\bar{\epsilon}~\Big[D^aD^b\Big(\frac{1}{T}\Big)-D^cD_c\Big(\frac{1}{T}\Big)h^{ab}\Big]\delta
h_{ab}\nonumber\\
&&=-\int_C\bar{\epsilon}~h^{a[b}h^{d]c}D_cD_d\Big(\frac{1}{T}\Big)\delta
h_{ab}\, ,
\end{eqnarray}
and
\begin{eqnarray}
\label{vE2}
&&\frac{1}{2}\int_C\bar{\epsilon}~\frac{\mathcal{B}_2}{T}=
\frac{1}{2}\int_C\bar{\epsilon}~\Big[2D^aD^b\Big(\frac{1}{T}\Big)R\nonumber\\
&&-2D^cD_c\Big(\frac{1}{T}\Big)Rh^{ab}-8D_cD_d\Big(\frac{1}{T}\Big)R^{ca}h^{db}\nonumber\\
&&+4D_cD_d\Big(\frac{1}{T}\Big)R^{ab}h^{cd}+4D_cD_d\Big(\frac{1}{T}\Big)R^{cd}h^{ab}\nonumber\\
&&+2D_cD_d\Big(\frac{1}{T}\Big)R^{adcb}-2D_cD_d\Big(\frac{1}{T}\Big)R^{acbd}\nonumber\\
&&-4D_c\Big(\frac{1}{T}\Big)h^{ab}\big(D^cR-2D_dR^{cd}\big)\nonumber\\
&&+4D^a\Big(\frac{1}{T}\Big)\big(D^bR-2D_cR^{cb}\big)\nonumber\\
&&-8D_c\Big(\frac{1}{T}\Big)\big(D^aR^{cb}-D^cR^{ab}+D_dR^{acbd}\big)\Big]\delta
h_{ab}\nonumber\\
&&=-\int_C\bar{\epsilon}~\Big[2Rh^{a[b}h^{d]c}D_cD_d\Big(\frac{1}{T}\Big)-4\big(R^{a[b}h^{c]d}\nonumber\\
&&-R^{c]b}h^{d[a}\big)D_cD_d\Big(\frac{1}{T}\Big)+2R^{acbd}D_cD_d\Big(\frac{1}{T}\Big)\Big]\delta
h_{ab}\, ,
\end{eqnarray}
where all of the terms with the derivatives of Riemann tensor are eliminated by using
Bianchi identity.

Remembering that  $T_0$ has been set to be a unit, so we
have  following relation
\begin{equation}
2\lambda^{-1}D_cD_d\lambda-\lambda^{-2}D_c\lambda
D_d\lambda=4TD_cD_d\Big(\frac{1}{T}\Big)\, .
\end{equation}
Thus the evolution equations become
\begin{equation}\label{MCT}
ph^{ab}=G^{ab}+\alpha H^{ab}
+2h^{a[b}h^{d]c}TD_cD_d\Big(\frac{1}{T}\Big)+4\alpha\Big[R^{a c b d}
-2 R^{a [b} h^{c]d} + 2R^{c[b}  h^{d]a}+R
h^{a[b}h^{d]c}\Big]TD_cD_d\Big(\frac{1}{T}\Big)\, .
\end{equation}
Combining Eqs. (\ref{vE}), (\ref{vE1}), (\ref{vE2}), and
(\ref{MCT}), we find that the variation of total entropy
({\ref{vS}}) is exactly vanishing, i.e., we have
\begin{equation}
\delta S=0\, .
\end{equation}
So far we have completed our proof of Theorem \ref{theorem1}.

\section{Maximum Entropy Principle in Lovelock Gravity}
In the previous sections, we have proved our Theorem \ref{theorem1}
in the context of Einstein-Gauss-Bonnet gravity. However, the
theorem can also be generalized to a more general Lovelock gravity.
The Lovelock action in $n$-dimensional spacetime is given by
\begin{equation}
I=\frac{1}{2}\int \epsilon
\sum_{i=0}^{[n/2]}\alpha_{(i)}\mathscr{L}_{(i)} +
I_{\mathrm{matter}}\, ,
\end{equation}
where $\epsilon$ is still the volume element of the static spacetime $(M, g_{ab})$, and $I_{\mathrm{matter}}$ also represents the action of matter fields. The coefficients $\alpha_{(i)}$ are constants and $\mathscr{L}_{(i)}$ is defined as
\begin{equation}
\mathscr{L}_{(i)}=\frac{1}{2^i}\delta^{a_1\cdots a_ib_1\cdots b_i}_{c_1\cdots c_id_1\cdots d_i}\mathscr{R}_{a_1b_1}{}^{c_1d_1}\cdots \mathscr{R}_{a_ib_i}{}^{c_id_i}\, .
\end{equation}
Here, $\delta^{a_1\cdots a_ib_1\cdots b_i}_{c_1\cdots c_id_1\cdots
d_i}=(2i)!\delta^{[a_1}_{c_1}\cdots
\delta^{a_i}_{c_i}\delta^{b_1}_{d_1}\cdots \delta^{b_i]}_{d_i}$ is
generalized Kronecker delta symbol. Considering the variation with
respect to $g_{ab}$, one gets the equations of motion which havw a
form
\begin{equation}
\label{EOMlovlock}
\mathscr{G}_{ab}=\sum_{i=0}^{[n/2]}\alpha_{(i)}
\mathscr{G}^{(i)}_{ab}=T_{ab}\, ,
\end{equation}
where
\begin{eqnarray}
\mathscr{G}^{(i)}{}^a_b=-\frac{1}{2^{i+1}}\delta^{a a_1\cdots a_ib_1\cdots b_i}_{bc_1\cdots c_id_1\cdots d_i}\mathscr{R}_{a_1b_1}{}^{c_1d_1}\cdots \mathscr{R}_{a_ib_i}{}^{c_id_i}\, .
\end{eqnarray}
For the static spacetime $(M, g_{ab})$, by using Eq.(\ref{RabcdtoRabcd}), after lengthy and tedious calculation, we find
\begin{eqnarray}
\mathscr{L}_{(i)}
 =L_{(i)} + i~G^{(i-1)}_{ab}\big(2\lambda{}^{-1}D^aD^b\lambda-\lambda^{-2} D^a\lambda D^b\lambda\big)\, ,
  \end{eqnarray}
 and
 \begin{eqnarray}
\mathscr{G}^{(i)}_{ab}={G}^{(i)}_{ab}-\frac{1}{2}\lambda^{-1}
L_{(i)}K_aK_b +
E_{acbd}^{(i)}\big(2\lambda^{-1}D^cD^d\lambda-\lambda^{-2}D^c\lambda
D^d\lambda\big)\, ,
\end{eqnarray}
where $L_{(i)}$ is the $i$th Lovelock Lagrangian on the hypersurface
$(\Sigma, h_{ab})$ and has an expression
\begin{equation}
L_{(i)}=\frac{1}{2^i}\delta^{a_1\cdots a_ib_1\cdots b_i}_{c_1\cdots
c_id_1\cdots d_i}R_{a_1b_1}{}^{c_1d_1}\cdots R_{a_ib_i}{}^{c_id_i}\,
,
\end{equation}
and $G^{(i)}_{ab}$ is the generalized ``Einstein tensor" for the
$i$th Lovelock Lagrangian in the hypersurface, i.e.,
\begin{equation}
\label{Giab}
G^{(i)}{}^a_b=-\frac{1}{2^{i+1}}\delta^{a a_1\cdots a_ib_1\cdots
b_i}_{bc_1\cdots c_id_1\cdots d_i}R_{a_1b_1}{}^{c_1d_1}\cdots
R_{a_ib_i}{}^{c_id_i}\, .
\end{equation}
In Eq.(\ref{Giab}), the tensor $E_{acbd}^{(i)}$ has a symmetry of Riemann tensor and is defined as
\begin{equation}
E_{acbd}^{(i)}=\frac{1}{2}\frac{\partial {L}_{(i)}}{\partial
R^{[ac][bd]}}\, .
\end{equation}
Based on the above discussion, we find the left-hand side of
Eq.(\ref{EOMlovlock}) can be transformed into a form
\begin{eqnarray}
\mathscr{G}_{ab}=\sum_{i=0}^{[n/2]}\alpha_{(i)}
\Big[{G}^{(i)}_{ab}-\frac{1}{2}\lambda^{-1} L_{(i)}K_aK_b +
E_{acbd}^{(i)}\big(2\lambda^{-1}D^cD^d\lambda-\lambda^{-2}D^c\lambda
D^d\lambda\big)\Big]\, .
\end{eqnarray}
This suggests that the Hamiltonian constraint in this theory can be written as
\begin{equation}
\rho=\frac{1}{2}\sum_{i=0}^{[n/2]}\alpha_{(i)} L_{(i)} \, .
\end{equation}
Similarly, the evolution equations are given by
\begin{eqnarray}
p~ h^{ab}=\sum_{i=0}^{[n/2]}\alpha_{(i)} \Big[{G}^{(i)ab}+
E^{(i)acbd}\big(2\lambda^{-1}D_cD_d\lambda-\lambda^{-2}D_c\lambda
D_d\lambda\big)\Big]\, .
\end{eqnarray}
With the preparation in the above, we can discuss the entropy
principle in this theory now. The variation of the total entropy
inside the region $C$ is still described by Eq.(\ref{vS}). We will
prove the Theorem \ref{theorem1} in the Lovelock gravity theory
order by order. We define
\begin{equation}\label{Ei}
\rho_i=\frac{1}{2}L_{(i)} \, ,
\end{equation}
and
\begin{equation}\label{Pi}
p_i~ h^{ab}={G}^{(i)ab}+
E^{(i)acbd}\big(2\lambda^{-1}D_cD_d\lambda-\lambda^{-2}D_c\lambda
D_d\lambda\big)\, .
\end{equation}
It should be noted here: $\rho_i$ and $p_i$ are not the real energy density and pressure of the fluid (the energy density and pressure of the fluid are denoted by $\rho$ and $p$ respectively). Actually, we have
\begin{equation}
\rho=\sum_{i}\alpha_{(i)}\rho_i\, ,\qquad p=\sum_{i}\alpha_{(i)} p_i\, .
\end{equation}
Consequently, the variation of the total entropy inside $C$ can be expressed as
\begin{equation}
\delta S=\sum_{i}\alpha_{(i)}\delta S_i\, ,
\end{equation}
where $\delta S_i$ is defined as
\begin{equation}\label{vSLi}
\delta S_i=\int_C\Big[\bar{\epsilon}~\frac{p_i}{2T}h^{ab}\delta
h_{ab}+\frac{1}{T}\delta(\bar{\epsilon}~\rho_i)\Big]\, .
\end{equation}
The second term in the integrand of the right-hand side of
Eq.(\ref{vSLi}) can be expanded as
\begin{equation}\label{vEi}
\int_C\frac{1}{T}\delta\big(\bar{\epsilon}~\rho_i\big)=\frac{1}{2}\int_C\frac{1}{T}\delta\big(\bar{\epsilon}~L_i\big)
=\frac{1}{2}\int_C\bar{\epsilon}~\frac{1}{T}\big(-G^{(i)ab}\delta
h_{ab}+\mathcal{B}_i\big)\, ,
\end{equation}
where $\mathcal{B}_i$ is the total derivative term which come from the variation of $L_i$
on the hypersurface.  Explicitly, this term can be written as
\begin{equation}
\mathcal{B}_i=2D_c\Big(\frac{\partial L_i}{\partial
R_{acbd}}D_b\delta h_{ad}\Big)-2D_b\Big(D_c\frac{\partial
L_i}{\partial R_{acbd}}\delta h_{ad}\Big)\, .
\end{equation}
By the same token in the Einstein-Gauss-Bonnet gravity, we can
rewrite the last term of Eq.(\ref{vEi}) as follows
\begin{eqnarray}\label{vEi1}
&&\frac{1}{2}\int_C\bar{\epsilon}~\frac{\mathcal{B}_i}{T}=\int_C\bar{\epsilon}~\frac{1}{T}\Big[D_c\Big(\frac{\partial
L_i}{\partial R_{acbd}}D_b\delta
h_{ad}\Big)-D_b\Big(D_c\frac{\partial L_i}{\partial R_{acbd}}\delta
h_{ad}\Big)\Big]\nonumber\\
&&=-\int_C\bar{\epsilon}~\Big[D_cD_d\Big(\frac{1}{T}\Big)\frac{\partial
L_i}{\partial R_{acbd}}\delta
h_{ab}+D_c\Big(\frac{1}{T}\Big)D_d\frac{\partial L_i}{\partial
R_{acbd}}\delta h_{ab}\Big]\nonumber\\
&&=-\int_C\bar{\epsilon}~D_cD_d\Big(\frac{1}{T}\Big)E^{(i)acbd}\delta h_{ab}\, ,
\end{eqnarray}
where we have used an identity $D_d(\partial L_i/{\partial
R_{acbd}})=0$ which can be deduced from Bianchi identity of Riemann
tensor. Actually, it is easy to find
\begin{eqnarray}
&&D_d\frac{\partial L_i}{\partial R_{acbd}}=D_d\Big(\frac{\partial
L_i}{\partial R_{ac}{}^{ef}}h^{eb}h^{fd}\Big)\nonumber\\
&&=h^{eb}D^f\Big(\frac{i}{2^i}\delta^{aca_1\cdots a_{i-1}c_1\cdots
c_{i-1}}_{efe_1\cdots e_{i-1}f_1\cdots
f_{i-1}}R_{a_1c_1}{}^{e_1f_1}\cdots
R_{a_{i-1}c_{i-1}}{}^{e_{i-1}f_{i-1}}\Big)\nonumber\\
&&=\frac{i}{2^i}h^{eb}\delta^{aca_1\cdots a_{i-1}c_1\cdots
c_{i-1}}_{efe_1\cdots e_{i-1}f_1\cdots
f_{i-1}}D^{[f}\Big(R_{a_1c_1}{}^{e_1f_1}\cdots
R_{a_{i-1}c_{i-1}}{}^{e_{i-1}f_{i-1}]}\Big)\nonumber\\
&&=0\, .
\end{eqnarray}
Considering the Tolman's law, and Eqs.(\ref{Pi}), (\ref{vEi}), and
(\ref{vEi1}), we find  $\delta S_i=0$ holds for each $i$. The total entropy thus
takes an extremum value, i.e.,
\begin{equation}
\delta S=\sum_{i=0}^{[n/2]}\alpha_i\delta S_i=0\, .
\end{equation}
By now, we complete our proof of the Theorem \ref{theorem1} in
Lovelock gravity.

In the above procedure, we can also reverse the proof by assuming
that total entropy already achieves extremum, and then the evolution
equations can be deduced from the Hamiltonian constraint. This can
be easily seen from Eq.(\ref{vSLi}). Thus we arrive at a converse
theorem
\begin{The}
 - Consider a perfect fluid in a static $n$-dimensional spacetime
$(M,g_{ab})$ in Lovelock gravity and $\Sigma$ is an
$(n-1)$-dimensional hypersurface orthogonal to the static Killing
vector field. Let $C$ be a region on $\Sigma$ with a boundary
$\partial{C}$. Assume that the temperature of the fluid obeys
Tolman's law and  both Hamiltonian constraint equation and fluid
equations are satisfied in $C$. Then the evolution equations are
implied by the extrema of the total fluid entropy for a fixed total
particle number in $C$ and for all variations in which  $h_{ab}$ and
its first derivatives are fixed on $\partial{C}$\, .
\end{The}

\section{spacetime with an $(n-2)$-dimensional maximally symmetric space}
With the assumption of maximally symmetric $(n-2)$-dimensional
space, our previous work~\cite{Cao:2013xy} has shown that the
generalized Tolman-Oppenheimer-Volkoff (TOV) equation can be deduced
from the (maximum) entropy principle together with the Hamiltonian
constraint in Lovelock gravity theory.

According to the present discussion, the (maximum) entropy principle
of perfect fluid can be realized  by using of EOM in such spacetime
manifestly. The $n$-dimensional spacetime metric is assumed as
\begin{equation}\label{maxmetr}
ds^{2}=-e^{2\Phi(r)}dt^{2}+e^{2\Psi(r)}dr^{2}+r^{2}\gamma_{ij}dz^{i}dz^{j}
\, ,
\end{equation}
where $\gamma_{ij}dz^idz^j$ is the metric of an $(n-2)$-dimensional
maximally symmetric space. The nontrivial components of the Riemann
tensor of such spacetime are given by
\begin{eqnarray}
&&R^{tr}_{~~tr}=-2e^{-2\Psi}(\Phi'^2-\Phi'\Psi'+\Phi'')\, ,\qquad R^{ti}_{~~tj}=\frac{e^{-2\Psi}}{r}\Phi'\delta^i_{~j}\, ,\nonumber\\
&&R^{ij}_{~~kl}=\frac{k-e^{-2\Psi}}{r^2}\delta^i_{~k}\delta^{j}_{~l}\,
,\qquad\qquad
R^{ri}_{~~rj}=-\frac{e^{-2\Psi}}{r}\Psi'\delta^i_{~j}\, .
\end{eqnarray}
where $k=0,\pm 1$ corresponds to the sectional curvature of the
maximally symmetric space and the prime denotes the derivative with
respect to radial coordinate $r$.The gravitational equations with
perfect fluid can be put in the form
\begin{equation}\label{spenergy}
\kappa_n^2 \rho=
\frac{1}{r^{n-2}}\frac{d}{dr}\Bigg{\{}\sum_{i=0}^{[n/2]}\frac{\alpha_i(n-2)!}{2(n-2i-1)!}r^{n-1-2i}\left(
k- e^{-2\Psi} \right)^{i}\Bigg{\}}\, ,
\end{equation}
which comes from $\mathcal{G}_{~t}^{t}=\kappa_n^2 T_{~t}^{t}$, and
\begin{eqnarray}\label{sppressure}
&&\kappa_n^2 p = \sum_{i=0}^{[n/2]}\frac{i\alpha_i(n-2)!}{(n-2i-1)!} \frac{e^{-2\Psi}\Phi' }{r}\left(  \frac{k-e^{-2\Psi} }{r^2} \right)^{i-1} \nonumber \\
&&- \sum_{i=0}^{[n/2]}\frac{\alpha_i(n-2)!}{2(n-2i-2)!}\left(
\frac{k- e^{-2\Psi}}{r^2} \right)^{i}\, ,
\end{eqnarray}
which is given by $\mathcal{G}_{~r}^{r}=\kappa_n^2 T_{~r}^{r}$.

The total entropy and total particle number inside $\partial C$ can
be written as
\begin{equation}
S=\omega_k\int_0^Rse^{\Psi}r^{n-2}dr\, ,
\end{equation}
\begin{equation}
N=\omega_k\int_0^Rne^{\Psi}r^{n-2}dr\, ,
\end{equation}
where $\omega_k:=\int d^{n-2}z \sqrt{\gamma}$ is the volume of the
maximally symmetric space with the sectional curvature $k$, and $R$
is the radius of spacial boundary $\partial C$

With additional assumptions of the Tolman's law (note that
$\sqrt{-\lambda}=e^{\Phi}$) and fixed particle number $N$ inside
$\partial C$, we can get that $\Phi'=-T'/T$ and the variation of
total entropy
\begin{equation}\label{delssym}
\delta
S=\omega_k\int_0^R\Big(\frac{p}{T}e^{\Psi}\delta\Psi+\frac{1}{T}\delta(e^{\Psi}\rho)\Big)r^{n-2}dr\,
.
\end{equation}
Substituting the gravitational equations Eqs.(\ref{spenergy}) and
({\ref{sppressure}}) into the above equation and performing the
integration by parts, we finally get $\delta S=0$ when the boundary
condition of $\Psi$: $\delta\Psi(R)=0$ is imposed.

Thus we realize the entropy principle in case of spacetime with
$(n-2)$-dimensional maximally symmetric space. We used the boundary
condition $\delta\Psi(R)=0$ in our derivation. For the metric of
Eq.(\ref{maxmetr}), the Misner-Sharp energy takes the
form~\cite{Hayward:1994bu, Misner:1964je}
\begin{equation}
\label{MSlovelock}
m(r)=\frac{\omega_k}{2\kappa_n^2}\sum_{i=0}^{[n/2]}\frac{\alpha_i(n-2)!}{(n-1-2i)!}r^{n-1-2i}\left(
k- e^{-2\Psi} \right)^{i}\, .
\end{equation}
Clearly, the boundary condition here has its physical meaning of
fixed Misner-Sharp energy inside the spacial boundary. The volume of
spacial boundary $A(R)=\int R^{n-2}\sqrt{\gamma}d^{n-2}z$ is held
fixed as well as the total particle number $N$ inside $\partial C$.
So $\delta m(R)=0$, $\delta A(R)=0$, and $\delta N=0$ define an
isolated system quasilocally.

In addition, we can reverse the procedure to get the so-called
generalized TOV equation by assuming that both the Hamiltonian
constraint \{here is the $tt$ component of gravitational equations
Eq.(\ref{spenergy})\} and equilibrium state of perfect fluid have
been already satisfied. Starting from Eq.(\ref{delssym}) with
$\delta S=0$ and the exact expression of $\rho$, i.e.,
Eq.(\ref{spenergy}), we obtain the evolution equation as
\begin{eqnarray}\label{TOV1}
&&\kappa_n^2 p = -\frac{T'}{T}\sum_{i=0}^{[n/2]}\frac{i\alpha_i(n-2)!}{(n-2i-1)!} \frac{e^{-2\Psi}}{r}\left(  \frac{k-e^{-2\Psi} }{r^2} \right)^{i-1} \nonumber \\
&&- \sum_{i=0}^{[n/2]}\frac{\alpha_i(n-2)!}{2(n-2i-2)!}\left(
\frac{k- e^{-2\Psi}}{r^2} \right)^{i}\, .
\end{eqnarray}
Note that the thermodynamic first law in terms of densities
Eq.(\ref{fl}) and Gibbs-Duhan relation Eq.(\ref{GD}) tells us that
\begin{equation}
dp=sdT+nd\mu \, ,
\end{equation}
together with Eq.(\ref{const}) we can get the following relation
\begin{equation}
\frac{T'}{T}=\frac{p'}{\rho+p}\, .
\end{equation}
So the evolution equation can be written as a relation among energy
density, pressure, and metric components. This is the so-called
generalized TOV equation in Lovelock gravity
\begin{eqnarray}\label{TOV2}
&&\kappa_n^2 p = -\frac{p'}{\rho+p}\sum_{i=0}^{[n/2]}\frac{i\alpha_i(n-2)!}{(n-2i-1)!} \frac{e^{-2\Psi}}{r}\left(  \frac{k-e^{-2\Psi} }{r^2} \right)^{i-1} \nonumber \\
&&- \sum_{i=0}^{[n/2]}\frac{\alpha_i(n-2)!}{2(n-2i-2)!}\left(
\frac{k- e^{-2\Psi}}{r^2} \right)^{i}\, .
\end{eqnarray}
More compactly, the generalized TOV equation can be written as
\begin{equation}
\frac{\partial
m}{\partial\Psi}\frac{p'}{\rho+p}=-\Big[p\omega_kr^{n-2}+m'-\frac{\partial
m}{\partial\Psi}\Psi'\Big]\, ,
\end{equation}
where $m(\Psi)$ is the Misner-Sharp energy Eq.(\ref{MSlovelock}) and
here it is understood as a function of $\Psi$.

As a little worm-up, we can see that at least in the case of
spacetime with maximally symmetric space, there do exist the
equivalent description between the geometrical equations and the
laws of thermodynamics without any man-made input since the the
boundary condition can be finally realized as the necessary
condition for an isolated system quasilocally.

The next section will focus on the static case, and talk about the
boundary conditions and its physical meaning.

\section{The boundary conditions and quasilocally isolated system}
In the previous sections, we have proved that the total entropy
inside the spacelike $(n-2)$-dimensional surface $\partial C$ will
take extremal value once the Tolman's law and both gravitational and
fluid equations are held. To get this conclusion, we have assumed
that the total particle number $N$ inside the spacial region $C$ is
held fixed and the variations $\delta h_{ab}$ and $D_a\delta h_{bc}$
are both vanishing on $\partial C$. The meaning of the condition
$\delta N$ is straightforward-there is no effective matter
communication between the inside and outside of $\partial C$. But
the physical meaning of the conditions $\delta h_{ab}=D_a\delta
h_{bc}=0$ is still unclear so far.

For a thermodynamic system by usual matter, the total entropy will
take maximal value for equilibrium state when the system is
isolated. In this case, the total energy $E$, volume $V$, and
particle number $N$ of the system are all held fixed, i.e., we have
\begin{equation}
\delta E=0\, ,\qquad\delta V=0\, ,\qquad\delta N=0\, .
\end{equation}

However, when gravity is taken into account, that is for a
self-gravitating one, the phrase ``isolated system'' becomes
ambiguous since the gravitational interaction is a long range force.
In Einstein gravity theory, the entropy extremum of relativistic
self-bound fluid in stationary axisymmetric spacetime has been
studied by Katz and Manor~\cite{Katz:1975},  they have presented the
condition for global isolation with globally defined quantities such
as total mass energy and total angular momentum. For a quasilocal
system, the (maximum) entropy problem of self-gravitating matter has
been studied by many authors~\cite{wald,gaosijie}, together with our
previous paper~\cite{Cao:2013xy}, where all the discussions are
based on the assumption of spacetime with maximally symmetric space.
Furthermore the quasilocal isolation is realized by requiring that
the Misner-Sharp energy $m(R)$ inside a fixed radius $R$ does not
change under the variation. The condition of fixed particle number
$N(R)$ appears as a Lagrange multiplier. So we see that at least in
spherically symmetric spacetime, the quasilocal realization of
isolation requires fixed Misner-Sharp energy $m(R)$, total particle
number $N(R)$ and fixed spacial area $\omega_kR^{n-2}$.

In this section, we will give the definition of isolated system
quasilocally for a more general static spacetime in Lovelock gravity
theory which has implied by the boundary conditions mentioned above
that was used to proof the extremum of total entropy.

First of all, a quasilocal system should have finite spacial volume
or more rigorously, such system has spacial boundary. We will denote
the product of surface $\partial C$ with segments of timelike
Killing vector field $K^a$'s integral curve as a timelike boundary
of spacetime manifold $M$ as $^{(n-1)}B$ with a unit norm $n^a$. The
induced metric and extrinsic curvature tensor of this timelike
boundary $^{(n-1)}B$ denoted as $\gamma_{ab}$ and $\Theta_{ab}$ have
the following form
\begin{equation}
\gamma_{ab}=g_{ab}-n_an_b\,
,\qquad\Theta_{ab}=\gamma_a{}^c\gamma_b{}^d\nabla_cn_d\, .
\end{equation}
Once a timelike boundary is imposed, the well-defined variational
principle requires a boundary term to cancel the total derivatives
that produce surface integrals involving the derivative of $\delta
g_{ab}$ normal to the boundary $^{(n-1)}B$. For The Loveloock
gravity theory, the boundary term can be written
as~\cite{Myers,Dehghani:2006ws}
\begin{equation}
I_b=-\int\tilde{\epsilon}\sum_{i=0}^{[n/2]}\sum_{s=0}^{i-1}\frac{(-)^{i-s}i\alpha_i}{2^s(2i-2s-1)}\mathcal{H}^{(i)}\,
,
\end{equation}
where $\tilde{\epsilon}$ is the volume element of $^{(n-1)}B$
associated with the induced metric $\gamma_{ab}$ and
$\mathcal{H}^{(i)}$ is defined as
\begin{equation}
\mathcal{H}^{(i)}=\delta_{b_1\cdots b_{2i-1}}^{a_1\cdots
a_{2i-1}}\mathscr{R}^{b_1b_2}{}{}_{a_1a_2}\cdots\mathscr{R}^{b_{2s-1}b_{2s}}{}{}_{a_{2s-1}a_{2s}}\Theta^{b_{2s+1}}_{a_{2s+1}}\cdots\Theta^{b_{2i-1}}_{a_{2i-1}}\,
.
\end{equation}
At this point,together with the boundary term, we have the total
action for a quasilocal gravitational system in Lovelock theory. It
is now straightforward to show that one can use the generalized
Hamilton-Jacobi method~\cite{Brown:1992br} to construct a divergence
free quasilocal stress-energy tensor defined on the timelike
boundary $^{(n-1)}B$ as~\cite{Myers,Dehghani:2006ws}
\begin{equation}\label{BYTab}
T^a_b=-\sum_{i=0}^{[n/2]}\sum_{s=0}^{i-1}\frac{(-)^{i-s}i\alpha_i}{2^s(2i-2s-1)}\mathcal{H}_b^{(i,s)a}\,
,
\end{equation}
where $\mathcal{H}_b^{(i,s)a}$ is
\begin{equation}
\mathcal{H}_b^{(i,s)a}=\delta_{b_1\cdots b_{2i-1}b}^{a_1\cdots
a_{2i-1}a}\tilde{R}^{b_1b_2}{}{}_{a_1a_2}\cdots\tilde{R}^{b_{2s-1}b_{2s}}{}{}_{a_{2s-1}a_{2s}}\Theta^{b_{2s+1}}_{a_{2s+1}}\cdots\Theta^{b_{2i-1}}_{a_{2i-1}}\,
,
\end{equation}
and $\tilde{R}_{abcd}$ is the intrinsic curvature tensor of the
timelike boundary $(^{(n-1)}B,\gamma_{ab})$.

Since the spacetime is static, the extrinsic curvature of spacelike
hypersurface is vanishing. So we can decompose the intrinsic and
extrinsic curvature tensor of $^{(n-1)}B$ along the time slice as
follows
\begin{equation}
\tilde{R}_{abcd}=\hat{R}_{abcd}-4u_{[a}\frac{\hat{D}_{b]}\hat{D}_{[c}\sqrt{-\lambda}}{\sqrt{-\lambda}}u_{d]}\,
,
\end{equation}
\begin{equation}
\Theta^a_b=k^a_b+u^au_bn_c\mathrm{a}^c\, ,
\end{equation}
where $\hat{R}_{abcd}$ is the intrinsic curvature tensor of
$(n-2)$-dimensional surface $(\partial C,\sigma_{ab})$ with the
covariant derivative operator denoted as $\hat{D}_a$ which can be
viewed as a submanifold embedded in spacelike hypersurface $\Sigma$
with a unit norm $n^a$. $\sigma_{ab}=h_{ab}-n_an_b$ and
$k_{ab}=-\sigma_a^cD_cn_b$ are induced metric and extrinsic
curvature tensor of $\partial C$'s , and $\mathrm{a}^{b}$ is the
acceleration vector of the static observer.

With the form of stress-energy tensor defined in Eq.(\ref{BYTab}),
one can obtain the energy density observed by the static observer as
\begin{equation}\label{qledensity}
\varepsilon=T^a_bu_au^b=-\sum_{i=0}^{[n/2]}\sum_{s=0}^{i-1}\frac{(-)^{i-s}i\alpha_{i}}{2^s(2i-2s-1)}t^{(i,s)}\,
,
\end{equation}
where
\begin{equation}\label{qledensityt}
t^{(i,s)}=\delta_{b_1\cdots b_{2i-1}b}^{a_1\cdots
a_{2i-1}a}u_au^b\hat{R}^{b_1b_2}{}{}_{a_1a_2}\cdots\hat{R}^{b_{2s-1}b_{2s}}{}{}_{a_{2s-1}a_{2s}}k^{b_{2s+1}}_{a_{2s+1}}\cdots
k^{b_{2i-1}}_{a_{2i-1}}\, ,
\end{equation}
which is the total energy density of the region $C$ quasilocally
defined on its spacial boundary $\partial C$. So the total energy of
the region $C$ as a thermodynamic quantity can be written as the
following integration
\begin{equation}\label{qlenergy}
E=\int\hat{\epsilon}\varepsilon\, ,
\end{equation}
where $\hat{\epsilon}$ is the volume element of $\partial C$.

Now let us study the physical implication of boundary conditions we
have imposed on variation of spacelike hypersurface $\Sigma$'s
induced metric $h_{ab}$ and its first derivatives. First, we noted
that the static observer's four velocity $u^a$ is hypersurface
orthogonal to $\Sigma$. This means that the (1,1)-type tensor field
$h^a_b$ is a projection operator which equals to a Kronecker delta
symbol of $\Sigma$ when restricted on the hypersurface $\Sigma$.
Then we can conclude that $\delta h^{ab}$ also vanishes on the
spatial boundary $\partial C$ because we can use the relation
$h^{ab}h_{bc}=h^a_c$ to deduce that
\begin{equation}
\delta h^{ab}=h^{bc}\delta h^a_c-h^{ac}h^{bd}\delta
h_{cd}=-h^{ac}h^{bd}\delta h_{cd}\, ,
\end{equation}
where the variation of $h^a_b$ vanishes since the variation is
restricted on $\Sigma$. Second, we find the following variational
relation by using the fact that $n^a$ is the unit norm of the
surface $\partial C$ embedded in spacelike hypersurface $\Sigma$
\begin{equation}
\delta n_a=\frac{1}{2}n_an^bn^c\delta h_{bc}\, .
\end{equation}
The (1-1)-type tensor $\sigma^a_b$ can also be viewed as a
projection operator of the surface $\partial C$ and equals  the
Kronecker delta symbol when restricted on it. Thus, the above
relation tells us that the variation of induced metric $\sigma_{ab}$
of $\partial C$ together with its all possible index form will
vanish when restricted on the surface $\partial C$.

Based on the above discussion, the variation of quasilocal energy
Eq.(\ref{qlenergy}) yields the following form
\begin{eqnarray}
&&\delta
E=\int\delta(\hat{\epsilon})\varepsilon+\int\hat{\epsilon}\delta(\varepsilon)\nonumber\\
&&=\int\frac{1}{2}\hat{\epsilon}\varepsilon\sigma^{ab}\delta\sigma_{ab}\nonumber\\
&&-\int\hat{\epsilon}\sum_{i=0}^{[n/2]}\sum_{s=0}^{i-1}\frac{(-)^{i-s}i\alpha_{i}}{2^s(2i-2s-1)}\Big[\delta^{a_1\cdots
a_{2i-1}a}_{b_1\cdots
b_{2i-1}b}\delta(u_au^b)\hat{R}^{b_1b_2}{}{}_{a_1a_2}\cdots\hat{R}^{b_{2s-1}b_{2s}}{}{}_{a_{2s-1}a_{2s}}k^{b_{2s-1}}_{a_{2s-1}}\cdots k^{b_{2i-1}}_{a_{2i-1}}\nonumber\\
&&+s\delta^{a_1\cdots a_{2i-1}a}_{b_1\cdots
b_{2i-1}b}u_au^b\hat{R}^{b_1b_2}{}{}_{a_1a_2}\cdots\delta\hat{R}^{b_{2s-1}b_{2s}}{}{}_{a_{2s-1}a_{2s}}k^{b_{2s-1}}_{a_{2s-1}}\cdots k^{b_{2i-1}}_{a_{2i-1}}\nonumber\\
&&+(2i-2s-1)\delta^{a_1\cdots a_{2i-1}a}_{b_1\cdots
b_{2i-1}b}u_au^b\hat{R}^{b_1b_2}{}{}_{a_1a_2}\cdots\hat{R}^{b_{2s-1}b_{2s}}{}{}_{a_{2s-1}a_{2s}}k^{b_{2s-1}}_{a_{2s-1}}\cdots\delta
k^{b_{2i-1}}_{a_{2i-1}}\Big]\, ,
\end{eqnarray}
where
\begin{equation}
\delta(u_au^b)=\delta h^b_a\, ,
\end{equation}
\begin{equation}
\delta\hat{R}^{ab}{}{}_{cd}=\sigma^{ae}\sigma^{bf}\sigma_{dg}\hat{D}_f(\sigma^{gm}\hat{D}_e\delta\sigma_{cm}+\sigma^{gm}\hat{D}_c\delta\sigma_{em}-\sigma^{gm}\hat{D}_m\delta\sigma_{ec})+2\hat{R}_e{}^b{}_{cd}\delta\sigma^{ae}+\hat{R}^{ab}{}{}_c{}^g\delta\sigma_{dg}\,
,
\end{equation}
\begin{equation}
\delta k^a_b=-\frac{1}{2}\sigma_b^ch^{ae}(D_c\delta h_{de}+D_d\delta
h_{ce}-D_e\delta h_{cd})-D_cn^a\delta\sigma_b^c-\sigma_b^cD_c\delta
n^a\, .
\end{equation}
Note the $\hat{D}_a$ is the covariant derivative operator which is
compatible with $\sigma_{ab}$ and $\partial C$ has no boundary. All
the variational terms in $\delta E$ can be finally changed into the
form which contains $\delta h_{ab}$ and $D_a\delta h_{bc}$. After
considering the integration on $\partial C$, the boundary conditions
stated in the theorems finally yield
\begin{equation}
\delta E=0\, ,
\end{equation}
which is nothing but the physical requirement that no energy
exchange with environment of an isolated system.

Since the total energy of the region $C$ is defined quasilocally, a
natural choice of volume of such a system is now the surface area of
the region $C$, that is the volume of the $(n-2)$-dimensional
surface $\partial C$
\begin{equation}
A=\int\hat{\epsilon}\, .
\end{equation}
According to our boundary conditions, it is easy to see that the
variation of this volume vanishes, i.e., $\delta A=0$.

Comparing our results with the  thermodynamic isolated system by the
usual matter, we can now claim that the boundary conditions together
with the fixed total particle number $N$ stated in theorems imply an
isolated system quasilocally in Lovelock gravity theory with
\begin{equation}
\delta E=0\, ,\qquad\delta A=0\, ,\qquad\delta N=0\, .
\end{equation}

As we have seen, the two boundary conditions $\delta h_{ab}=0$ and
$D_a\delta h_{bc}=0$ on the spacelike $(n-2)$-dimensional surface
$\partial C$ are necessary conditions to define a quasilocally
isolated system in Lovelock gravity theory.

If we relax one of the boundary conditions, then we can expect to
get the variational relation among the total entropy and other
thermodynamic quantities. Let us take Einstein gravity as a simple
example. We will relax the boundary condition $D_a\delta h_{bc}=0$
on $\partial C$ and see the result of entropy variation.

Without the assumption of vanishing of the first derivative of all
variations of induced metric on $\partial C$, the variation of total
entropy now is nonzero even if the equations of motion of gravity
are satisfied.
\begin{eqnarray}
\delta S&=&\int_C\Big[\bar{\epsilon}\frac{p}{2T}h^{ab}\delta
h_{ab}+\frac{1}{T}\delta(\bar{\epsilon}\rho)\Big]\nonumber\\
&=&\int_C\Big[\bar{\epsilon}\frac{p}{2T}h^{ab}\delta
h_{ab}+\frac{1}{2T}\delta(\bar{\epsilon}R)\Big]\nonumber\\
&=&\int_C\bar{\epsilon}\Big[h^{a[b}h^{d]c}D_cD_d\Big(\frac{1}{T}\Big)\delta
h_{ab}+\frac{1}{2T}D^aD^b\delta
h_{ab}-\frac{1}{2T}D^a(h^{bc}D_a\delta h_{bc})\Big]\nonumber\\
&=&\int_C\bar{\epsilon}\Big[D^a\Big(\frac{1}{2T}D^b\delta
h_{ab}\Big)-D^a\Big(\frac{1}{2T}h^{bc}D_a\delta
h_{bc}\Big)\Big]\nonumber\\
&=&\int_{\partial
C}\frac{\hat{\epsilon}}{T}\Big[\frac{1}{2}n^aD^b\delta
h_{ab}-\frac{1}{2}n^aD_a(h^{bc}\delta h_{bc})\Big]\nonumber\\
&=&-\frac{1}{2}\int_{\partial
C}\frac{\hat{\epsilon}}{T}n^aD_a(h^{bc}\delta h_{bc})\, ,
\end{eqnarray}
In the last step, we have used the fact that $D^b\delta
h_{ab}=\sigma^{be}D_e\delta h_{bd}=0$ because $\delta h_{ab}=0$ on
$\partial C$.

On the other hand, one can define the quasilocal energy inside
$\partial C$ according to Eqs.(\ref{qledensity}),
(\ref{qledensityt}), and (\ref{qlenergy}) when limited in the
Einstein gravity case as
\begin{equation}
E=-\int_{\partial C}\hat{\epsilon}\sigma^a{}_bD_an^b\, .
\end{equation}
If we vary the above energy only with $\delta h_{ab}=0$ on $\partial
C$, then we will find
\begin{eqnarray}
\delta E&=&-\int_{\partial C}\hat{\epsilon}h^a{}_b\delta
C^b{}_{ac}n^c=-\frac{1}{2}\int_{\partial
C}\hat{\epsilon}h^a{}_bh^{bd}(D_a\delta h_{cd}+D_c\delta
h_{ad}-D_d\delta h_{ac})n^c\nonumber\\
&=&-\frac{1}{2}\int_{\partial C}\hat{\epsilon}n^aD_a(h^{bc}\delta
h_{bc})\, .
\end{eqnarray}
Thus, when $\partial C$ is an isothermal boundary, the variation of
total entropy inside $\partial C$ can be written as
\begin{equation}
\delta S=\frac{1}{T}\delta E\, .
\end{equation}
This is nothing but the thermodynamic first law of the isometrical
system since we have fixed the induced metric on $\partial C$.

\section{Conclusions and Discussion}
In this paper, we have shown that the (maximum) entropy principle of
the perfect fluid in curved spacetimes can be realized by using the
gravitational equations  in the Lovelock gravity theory. This result
has been put into the Theorem \ref{theorem1}. Comparing to our
previous paper, the symmetry of an $(n-2)$-dimensional maximally
symmetric space has not been imposed, and the only symmetry required
here is the static condition.

For the traditional thermodynamics in flat spacetime, the entropy of
matter must take maximal value in an equilibrium state if the system
is isolated. When backreaction is encountered, that is for
self-gravitating system, it seems that the requirement for isolation
at least includes following conditions: First, the system inside an
$(n-2)$-dimensional spacelike surface $\partial C$ should have a
fixed total particle number. Second,  the induced metric $h_{ab}$ on
$\Sigma$ and its first derivatives should be fixed on $\partial C$.
Physically, the first condition implies that the system has no
effective particle communication with the outside region. The second
one implies two physical explanations, one is the volume of the
system which is quasilocally defined as the spacial region $C$'s
surface area should keep fixed, the other is the total quasilocal
energy of the system does not change under the variations of the
matter fields. Thus these two conditions in Theorem 1 and 2 will
give the definition of an isolated system quasilocally when
backreaction is taken into account.

We have just shown that the total entropy of the perfect fluid for
this isolated system must take extremum value when both the
gravitational and fluid equations are satisfied. However, we do not
know the extremum is a maximum or a minimum  at present. To confirm
the state is a real equilibrium state, one has to perform a second
order variation and analyze the stability conditions of the system.
The maximum entropy principle in general relativity with stability
analysis in spherical symmetric system has been studied by
Roupas~\cite{Roupas:2013nt}. This is an interesting point and needs
further study in static spacetime. This is also the reason that we
have put the ``maximum'' inside brackets in the title and the main
part of this paper.

It is still unclear whether this (maximum) entropy principle can be
extended to other gravity theories or not. We believe this deep connection between gravity theory and thermodynamics
is still there and waiting for people to uncover.

Finally, is it possible to apply the (maximum) entropy principle to
a stationary spacetime? This is also unclear up to date and deserves
to be studied carefully in the future.

\section{Acknowledgments}

This work was supported in part by the National Natural Science
Foundation of China with Grants  No.11205148 and No.11235010. L. M.
C would like to thank Sijie Gao for his useful communication and
comment.



\begin{thebibliography}{99}


\bibitem{bardeen1973} J.~M.~Bardeen, B.~Carter and S.~W.~Hawking,
  Commun.\ Math.\ Phys.\  {\bf 31}, 161 (1973).

\bibitem{hawking}S.~W.~Hawking,
  Commun.\ Math.\ Phys.\  {\bf 43}, 199 (1975) \  {\bf 46}, 206E (1976)].

\bibitem{Bekenstein} J.~D.~Bekenstein,
Phys.\ Rev.\  D {\bf 7}, 2333 (1973).


\bibitem{Wald:1999vt}
  R.~M.~Wald,
  Living Rev.\ Relativity.\  {\bf 4}, 6 (2001)
  [arXiv:gr-qc/9912119].

\bibitem{Brown:1992br}
  J.~D.~Brown and J.~W.~York, Jr.,
  Phys.\ Rev.\ D {\bf 47}, 1407 (1993)
  [gr-qc/9209012].

\bibitem{Hayward:1994bu}
  S.~A.~Hayward,
  Phys.\ Rev.\ D {\bf 53}, 1938 (1996)  [gr-qc/9408002].  

\bibitem{Ashtekar:2004cn}
  A.~Ashtekar and B.~Krishnan,
  Living Rev.\ Relativity.\  {\bf 7}, 10 (2004)
  [gr-qc/0407042].

\bibitem{Jac}T.~Jacobson,
Phys.\ Rev.\ Lett.\  {\bf 75}, 1260 (1995) [arXiv:gr-qc/9504004].

\bibitem{Jac1}
  C.~Eling, R.~Guedens and T.~Jacobson,
  Phys.\ Rev.\ Lett.\  {\bf 96}, 121301 (2006)  [gr-qc/0602001].

\bibitem{c1}
  R.~G.~Cai and S.~P.~Kim,
  J. High Energy Phys. {\bf 02} (2005) 050  [hep-th/0501055].


\bibitem{wald}
  R.~D.~Sorkin, R.~M.~Wald and Z.~J.~Zhang,
  Gen.\ Relativ.\ Gravit.\  {\bf 13}, 1127 (1981).  

\bibitem{gaosijie}
  S.~Gao,
  Phys.\ Rev.\ D {\bf 84}, 104023 (2011); 85, 027503 (2012) [arXiv:1109.2804 [gr-qc]].

\bibitem{Anastopoulos:2013xdk}
  C.~Anastopoulos and N.~Savvidou,
  Classical Quantum Gravity {\bf 31}, 055003 (2014)
  [arXiv:1302.4407 [gr-qc]].

\bibitem{Fang:2013oka}
  X.~Fang and S.~Gao,
  Phys.\ Rev.\ D {\bf 90}, 044013 (2014)
  [arXiv:1311.6899 [gr-qc]].



\bibitem{Love}D.~Lovelock,
  J.\ Math.\ Phys.\  (N.Y.) {\bf 12}, 498 (1971).

\bibitem{Zwie}
 B.~Zwiebach,
  Phys.\ Lett.\ B {\bf 156}, 315 (1985).

\bibitem{Lowenergylimit}
  D.~J.~Gross and E.~Witten,
  Nucl.\ Phys.\  B {\bf 277}, 1 (1986).

 \bibitem{Lowenergylimit1}
  R.~R.~Metsaev and A.~A.~Tseytlin,
Nucl.\ Phys.\  B {\bf 293}, 385 (1987).

\bibitem{Lowenergylimit2}
I.~Jack and D.~R.~T.~Jones,
  Nucl.\ Phys.\  B {\bf 303}, 260 (1988).

\bibitem{Lowenergylimit3}
K.~A.~Meissner,
  Phys.\ Lett.\  B {\bf 392}, 298 (1997)
  [arXiv:hep-th/9610131].

\bibitem{Cao:2013xy}
  L.~M.~Cao, J.~Xu and Z.~Zeng,
  Phys.\ Rev.\ D {\bf 87}, 064005 (2013)
  [arXiv:1301.0895 [gr-qc]].


\bibitem{Green:2013ica}
  S.~R.~Green, J.~S.~Schiffrin, and R.~M.~Wald,
  Classical Quantum Gravity {\bf 31}, 035023 (2014)
  [arXiv:1309.0177 [gr-qc]].


\bibitem{Wald:1984rg}
  R.~M.~Wald,
  General Relativity (University Press, Chicago, 1984), p. 491.


\bibitem{Geroch:1970nt}
  R.~P.~Geroch,
  J.\ Math.\ Phys.\  (N.Y.) {\bf 12}, 918 (1971).


\bibitem{Misner:1964je}
  C.~W.~Misner and D.~H.~Sharp,
  Phys.\ Rev.\  {\bf 136}, B571 (1964).



\bibitem{Katz:1975}
 J.~Katz and Y.~Manor,
 Phys.\ Rev.\ D {\bf 12}, 956 (1975).

\bibitem{Myers}
 R.~C.~Myers,
 Phys.\ Rev.\ D {\bf 36}, 392 (1987).


\bibitem{Dehghani:2006ws}
  M.~H.~Dehghani, N.~Bostani and A.~Sheykhi,
  Phys.\ Rev.\ D {\bf 73}, 104013 (2006)
  [hep-th/0603058].


\bibitem{Roupas:2013nt}
  Z.~Roupas,
  Classical Quantum Gravity {\bf 30}, 115018 (2013)
  [arXiv:1301.3686 [gr-qc]].

\end{thebibliography}
\end{document}